\documentclass[aps,prl,twocolumn,showpacs,floatfix,superscriptaddress]{revtex4}
\usepackage{graphicx}

\begin{document}
\flushbottom

\title{Supercurrent on a vortex core in 2H-NbSe$_2$: current driven scanning tunneling spectroscopy}
\author{A. Maldonado}
\affiliation{Laboratorio de Bajas Temperaturas, Departamento de F\'isica de la Materia
Condensada, Instituto de Ciencia de Materiales Nicol\'as Cabrera and
Condensed Matter Physics Center, Universidad Aut\'onoma de Madrid, E-28049 Madrid,
Spain}
\author{S. Vieira}
\affiliation{Laboratorio de Bajas Temperaturas, Departamento de F\'isica de la Materia
Condensada, Instituto de Ciencia de Materiales Nicol\'as Cabrera and
Condensed Matter Physics Center, Universidad Aut\'onoma de Madrid, E-28049 Madrid,
Spain}
\affiliation{Unidad Asociada de Bajas Temperaturas y Altos Campos Magn\'eticos, UAM/CSIC, Cantoblanco, E-28049 Madrid, Spain}
\author{H. Suderow}
\email[Corresponding author: ]{hermann.suderow@uam.es}
\affiliation{Laboratorio de Bajas Temperaturas, Departamento de F\'isica de la Materia
Condensada, Instituto de Ciencia de Materiales Nicol\'as Cabrera and
Condensed Matter Physics Center, Universidad Aut\'onoma de Madrid, E-28049 Madrid,
Spain}
\affiliation{Unidad Asociada de Bajas Temperaturas y Altos Campos Magn\'eticos, UAM/CSIC, Cantoblanco, E-28049 Madrid, Spain}

\begin{abstract}
We report current driven scanning tunneling spectroscopy (CDSTS) measurements at very low temperatures on vortices in 2H-NbSe$_{2}$. We find that a current produces an increase of the density of states at the Fermi level in between vortices, and a reduction of the zero bias peak at the vortex center. This occurs well below the de-pairing current. We conclude that a supercurrent affects the low energy part of the superconducting gap structure of 2H-NbSe$_{2}$.
\end{abstract}

 \date{\today}
\maketitle

The nature of vortex cores in superconductors and superfluids has been matter of research for decades. At the core, the decay of the modulus of the pair wavefunction, occurring over the coherence length $\xi$, is continuous up to the center, where it vanishes. Caroli, de Gennes and Matricon demonstrated that vortex cores hold bound quasiparticle states inside\cite{Caroli64}. Such states were first visualized by H.F. Hess et al. using scanning tunneling microscopy in the superconductor 2H-NbSe$_2$\cite{Hess89,Hess90,Guillamon08,Guillamon08c,Fischer07,Nishimori04,Song11,Allan13,Zhou13}. Quasiparticles in a spatially varying pair potential acquire mixed electron-hole character. They suffer Andreev reflection when the modulus of the pair wavefunction gains its maximal value. In the vortex core of a clean superconductor (with mean free path $\ell\gg\xi$), Andreev reflection gives bound states which form when the phases of the multiply reflected quasiparticle wavefunctions interfere constructively. Caroli et al. found that the lowest energy state is located at $\Delta^{2}/2E_{F}$; i. e., effectively close to the Fermi level in many materials\cite{Caroli64}. The discovery of these core Andreev bound states led to new insight\cite{Gygi91,Hayashi98,Melnikov09}, and was used to explain macroscopic effects, such as the absence of thermal conductivity in some materials when the magnetic field is parallel to the heat gradient\cite{Boaknin03}. Andreev core states play a determinant role in explaining the onset of dissipation in moving vortices, because the scattering processes between Andreev states and the rest of the electronic system can produce dissipation, depending on energy level spacing and scattering rates\cite{Kopnin02,Kopnin00b,Kopnin96,Sonin13,Stone00,Stone96,Guinea95}. Here we address with current driven scanning tunneling spectroscopy (CDSTS \cite{Maldonado11}) how vortex bound states are affected by a current flow through the superconductor at low currents, far below the de-pairing current and when vortices are still static. We find a significant effect, and relate it to the multiband superconducting properties of 2H-NbSe$_2$\cite{Rodrigo04c,Yokoya01,Kiss07,Boaknin03,Guillamon08,Guillamon07,Hess91}.

\begin{figure}[ht]
\includegraphics[width=8cm,keepaspectratio, clip]{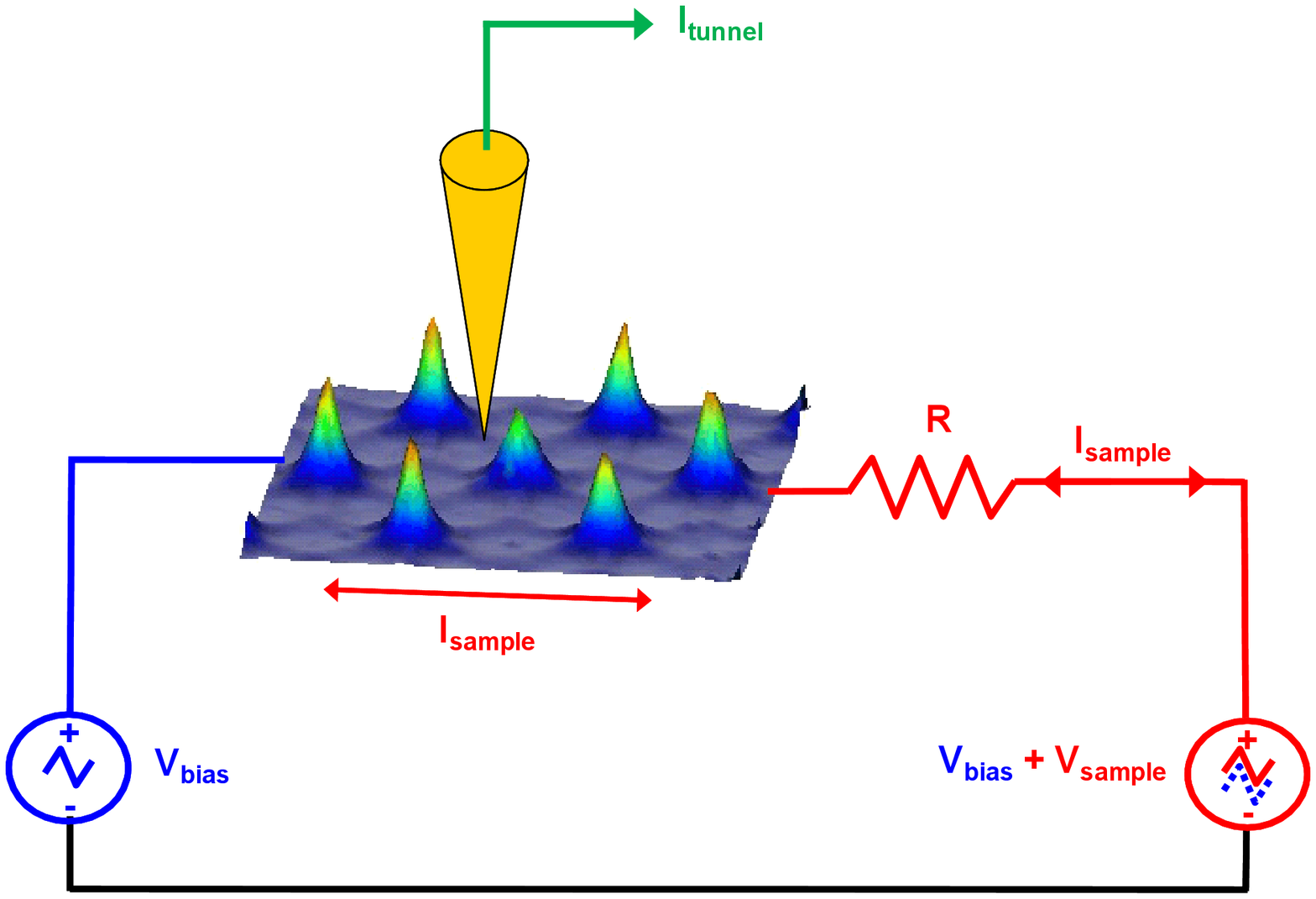}
\vskip -0cm \caption{(Color online) Schematics of the CDSTS experiment performed using a tip of Au and a sample of 2H-NbSe$_{2}$. A current I$_{sample}$ is applied through the sample. In order to maintain this current at a constant value ($V_{sample}/R$) when measuring the bias voltage dependence of the conductance, we add $V_{sample}$ to the bias voltage ramp V$_{bias}$. The tunneling current I$_{tunnel}$ is measured through the tip electrode using an I/V converter. The current is applied in the vortex-vortex nearest neighbor direction, as indicated by the red arrow. Details of the experimental procedure are given in Ref.\protect\cite{Maldonado11}.} \label{Fig1}
\end{figure}

We use a home-made STM thermally anchored to the mixing chamber of a dilution refrigerator equipped with a superconducting magnet. The magnetic field is always applied normal to the sample's surface. We apply a constant transport current through the sample $I_{sample}$ of 10.6 mA, maintaining the full conductance vs bias voltage capability of the STM/S experiment (Fig.\ 1 and Ref. \cite{Maldonado11}). We focus on measurements made at 0.1 T, where the vortex core states and star-shape are well resolved with zero current. The tunneling current is measured through the tip electrode as a function of the bias voltage without changing I$_{sample}$ by applying the ramp V$_{bias}$ on both sides of the sample. The tunneling conductance $g(V_{bias},I_{sample})$ is obtained through a numerical derivative dI$_{tunnel}$/dV$_{bias}$\cite{Suderow11,Guillamon08}. $g(V_{bias},I_{sample})$ is normalized to the value obtained at a bias voltage of 2.5 mV. The energy resolution of this system, as checked by measuring superconducting tunneling conductance using Al as the tip and the sample electrodes\cite{Rodrigo04}, is of 35 $\mu$eV\cite{Maldonado11}. This value, which does not vary when applying a current, corresponds to an effective temperature of 200 mK. We use tips of Au, which are prepared and cleaned in-situ as described in Ref. \cite{Rodrigo04b}, and a 2H-NbSe$_{2}$ single crystal grown by iodine vapour deposition, with dimensions of about 3 mm $\times$ 3 mm $\times$ 0.1 mm. The sample was glued on top of two silver epoxy contacts made over the full length of two sides of the sample. The resistance of the current circuit was dominated by the resistance of the copper wires used, which avoided Joule overheating at the sample due to the applied current. The sample was isolated from the sample holder, which was grounded, using a Kapton foil. It was mechanically exfoliated just before mounting it on the STM and cooling down. The tip was located in an area close to the center of the surface of the sample. The topography images obtained in this sample presented atomic resolution and charge density wave (CDW) order features as in previous work\cite{Guillamon08c}. The vortex lattice is oriented along one of the high symmetry axis of the hexagonal Se lattice\cite{Hess91,Hess90,Hess89,Guillamon08,Guillamon07,Guillamon08c}. Thus, we can determine the x-y scan direction with the orientation of the atomic and vortex lattices. The x-y scanner orientation with respect to the current leads is further known by optically inspecting the sample holder and piezo as mounted. From this, we determine that the current is applied parallel to the vortex-vortex nearest neighbor direction, as schematically shown in Fig.\ 1.

\begin{figure}[ht]
\includegraphics[width=8cm,keepaspectratio, clip]{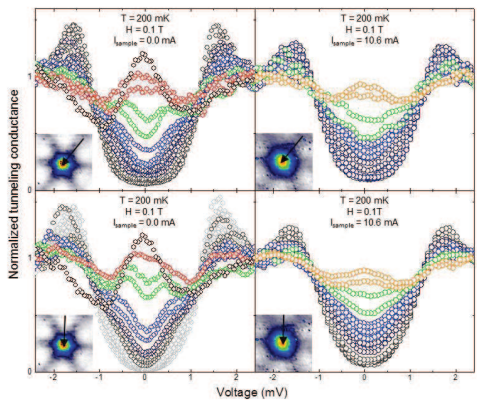}
\vskip -0cm \caption{(Color online) Normalized tunneling conductance curves as a function of the bias voltage obtained at 200 mK, under a magnetic field of 0.1 T and zero (left panels) and 10.6 mA of current applied through the sample (right panels), when approaching the center of a vortex. In the upper panels, the curves have been taken along the direction of a ray, whereas in the bottom panels, these have been taken along a direction in between rays. Notice that curves have been color-coded as in the zero bias conductance images shown in the insets.} \label{Fig2}
\end{figure}

Fig. 2 shows the comparison between the spatial variation of the tunneling conductance spectra when approaching the center of a vortex core without and with a $I_{sample}$ = 10.6 mA. As it is well known, at zero current, the zero bias Andreev bound states peak at the vortex center\cite{Caroli64} is displaced to higher energies when moving away from the center (Fig. 2(left panels)). There is an in-plane anisotropy observed in the tunneling conductance curves when the vortex center is approached\cite{Gygi91,Hayashi97}. The corresponding conductance maps at zero bias are shown in the insets of Fig.2, with a six-fold star shape whose rays are located at 30$^\circ$ with respect to the vortex lattice\cite{Hess91,Hess90,Hess89,Guillamon08,Guillamon07,Guillamon08c}. Maps at other bias voltages are shown in Fig.3. The star turns by $30^\circ$ at a bias voltage of 0.5 mV and the contrast inverts at bias voltages corresponding to the quasiparticle peaks \cite{Hess91,Hess90,Hess89,Guillamon08,Guillamon07,Guillamon08c}. Under the applied current, qualitatively the same features remain (Fig. 2(right panels) and Fig.\ 3(b)). However, the peak at zero bias at the center of the vortex core is reduced (Fig. 2(right panels)) and the spatial variations in conductance which give the six-fold star shape are decreased. The $30^\circ$ turning of the star is smeared out and the contrast inversion occurs at roughly the same bias voltage (Fig. 3(b)).

\begin{figure}[ht]
\includegraphics[width=8cm,keepaspectratio, clip]{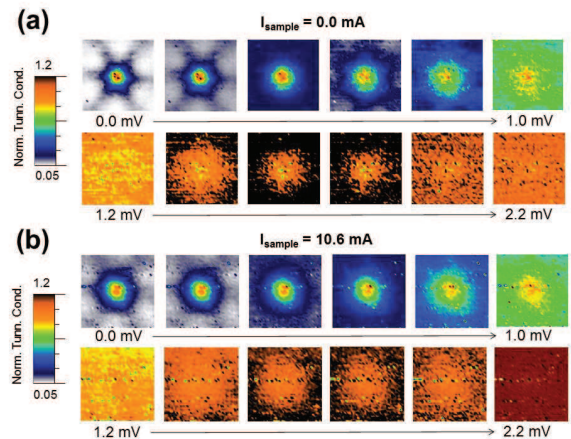}
\vskip -0cm \caption{(Color online) Evolution of the shape of the vortex core as a function of the bias voltage at zero current (a) and with 10.6mA (b). Color scale follows the code used in the full tunneling conductance curves shown in Fig.\ 2 and in the insets. Data are taken at 200 mK and at 0.1 T.} \label{Fig3}
\end{figure}

Results obtained at zero field are shown in Fig.\ 4(a). We observe an increase of the electronic density of states at low energies and a slight reduction of the height of the quasiparticle peaks.

It is useful to discuss our observations in relation to the de-pairing $J_d$ current. A simple, yet useful understanding of $J_d$ can be obtained by taking into account that the current Doppler-shifts the superconducting density of states\cite{Volovik93,Anthore03,Sanchez01,Suderow97,Suderow98,Suderow02,Kohen06}. At the de-pairing current density $J_d$ the Doppler shift is of order of the gap and superconductivity is lost. Unfortunately, the flow of a current through the cross section of a bulk superconducting sample is a difficult problem involving electrodynamics and vortex matter which is not fully solved\cite{Blatter94,Brandt01,Brandt93}. In a bulk type II superconductor at zero field the current density distribution is inhomogeneous over the cross section of the sample, in particular close to the edge of the sample\cite{Brandt01,Brandt93}. Geometrical shapes with vortex and Meissner phases can appear, whose distribution may be influenced by multiband superconducting properties\cite{Brandt01,Prozorov08,Moshchalkov09,Brandt10,Silaev12,Gutierrez12}. Under magnetic fields in the vortex phase, the magnetic field of each vortex overlaps when the intervortex distance $a_0$ drops below $\lambda$. For instance, in 2H-NbSe$_2$, $a_0\approx\lambda$ at 0.1 T (taking $\lambda$ of $0.15 \mu m$ \cite{Fletcher07}). As the magnetic field penetrates over the whole sample, one may assume that then the current distributes homogeneously over the thickness\cite{VanDuzer}. However, there is experimental evidence for a dependence of the current distribution on the pinning properties, even far below the de-pinning current. Neutron scattering in a PbIn polycrystal with weak pinning shows that the current distribution well below de-pinning is similar than at zero field\cite{Pautrat03}. 2H-NbSe$_2$ is a material with weak pinning\cite{Banerjee98,Paltiel00,Pasquini08,Pautrat04}. If we assume that the current flows at least over the magnetic penetration depth $\lambda$ close to the surface, we obtain a maximal value for the current density  $J_{sample}\approx 10^3 A/cm^2$ both at zero field and in the mixed state. $J_d$ can be estimated using $J_d$ = $\frac{H_c}{\lambda}$\cite{DewHughes01,Fridman11}, and we find, taking $B_c$ = 130 mT, with $\lambda$ = 0.15 $\mu$m and $\xi$ = 10 nm, $J_d \sim 10^{7} A/cm^2$. This value is four orders of magnitude above $J_{sample}$. Thus, the current density in our experiment is far below de-pairing $J_d$ value.

The superconducting density of states of vortex free areas at zero current is significantly different than the one found at zero field. Close to zero bias, the tunneling conductance is smeared (Fig.\ 4) and the quasiparticle peaks are slightly reduced with respect to the zero field result. This has been discussed previously in macroscopic experiments, and shows that the magnetic field affects the gap distribution over the Fermi surface \cite{Boaknin03}. Not only NbSe$_2$, but also other multiband superconductors, such as MgB$_{2}$\cite{Bouquet01,Sologubenko02}, CeCoIn$_{5}$\cite{Seyfarth08} and PrOs$_{4}$Sb$_{12}$\cite{Seyfarth05,Seyfarth06} show a strong increase of excitations when applying a magnetic field.

When we apply a current, there is an additional reduction of the superconducting density of states at the quasiparticle peaks, and an increase in the zero bias conductance in vortex free areas. Thus, the application of the current affects the band structure by further promoting the decrease of smaller sized gaps over the Fermi surface (Fig.\ 4b). Remarkably, this occurs with a current whose value is far below de-pairing current $J_d$ estimated above. The Fermi surface of 2H-NbSe$_2$ consists of nearly two-dimensional (2D) concentric cylinders, which derive from Nb 4d orbitals, and a single small three-dimensional (3D) pancakelike sheet derived from Se 4p orbitals, as inferred from angular resolved photoemission (ARPES) and de Haas-van Alphen experiments\cite{Corcoran94,Yokoya01,Kiss07,Johannes06,Suderow05d}. The gap size, as measured in Ref.\cite{Kiss07}, is found to be smallest along one direction in one Nb derived 2D sheet. In the central three dimensional sheet, the gap practically disappears when increasing temperature, as shown by ARPES and temperature dependent tunneling spectroscopy\cite{Rodrigo04c}. The anisotropy of the 2D sheets is significant, and has been associated to the CDW and the six-fold in-plane star shaped vortex core\cite{Hayashi97,Guillamon08}. Our data show that the six-fold star shape is essentially maintained under an applied current, although it is significantly smeared. Thus, it seems reasonable to conclude that the parts of the Fermi surface with smaller sized gap values, which are also affected by temperature\cite{Kiss07,Rodrigo04c}, are more sensitive than the rest of the Fermi surface to the application of a current. The states created in between vortices deplete the Andreev core levels and give a reduction of the zero bias peak found at the center of the vortex core.

It has been shown that Andreev bound states inside a vortex core are different from quantum levels inside a potential well in that they are not fully localized\cite{Bardeen72,Heida98,Sonin13,Rainer96}. Andreev reflection provides a smooth connection between core levels and the surrounding superconductor. Our results show that the destruction of the gap in between vortices due to the current leads to a decrease of the amount of Andreev levels in the core. The effect of the current is, in this sense, similar to a pair breaking effect due to impurities or dopants \cite{Renner91} or a magnetic field.

Following Refs.\cite{Kopnin02,Kopnin00b,Kopnin96,Sonin13,Stone00,Stone96}, vortex motion is non-dissipative when the quasiparticle scattering rate $1/\tau$ is well below core level separation. As we show here, the application of a current reduces core level separation, leading to a more continous-like core spectrum in superconductors with a structured gap over the Fermi surface. Thus, it seems easier to produce dissipation through vortex motion if the gap shows a strong Fermi surface dependence.

\begin{figure}[ht]
\includegraphics[width=8cm,keepaspectratio, clip]{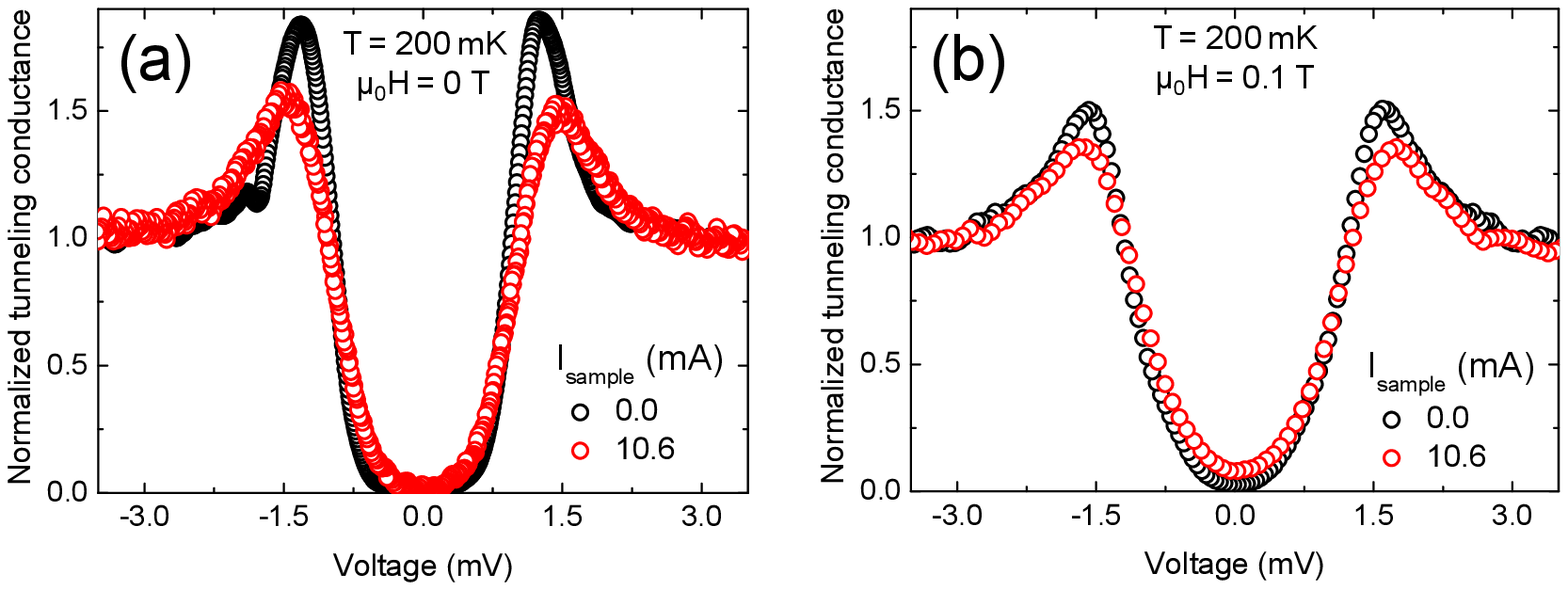}
\vskip -0cm \caption{(Color online) (a) Zero field normalized tunneling conductance curves taken at 200 mK. (b) Normalized tunneling conductance curves as a function of the bias voltage obtained at 200 mK and 0.1 T far away from the vortex and in between rays. In both cases, the black points correspond to the measurements performed without the application of any current and the red ones to the ones taken under a $I_{sample}$ = 10.6 mA.} \label{Fig4}
\end{figure}

In Fig.\ 5 we compare the spatial evolution of the zero bias normalized tunneling conductance $G(0 mV)$ with and without current flow along two characteristic directions of the core star-shape. The spatial variation of the conductance is reduced by the applied current in all directions, giving vortices, which, as measured with tunneling spectroscopy, are larger than without a current.

There have been several works discussing vortex core radius shrinking in 2H-NbSe$_2$ and other compounds when increasing the magnetic field, in particular from Muon scattering experiments\cite{Callaghan05,Sonier04,Kogan05}. The same experiments find a significant increase of the amount of quasiparticle states in between vortices with the magnetic field\cite{Sonier04}. It is proposed that different Andreev levels are formed in vortex cores of multiband superconductors, associated to different parts of the Fermi surface with varying gap size. The magnetic field induced disappearance of the smaller sized superconducting gaps depletes core states created by smaller gap. This, in turn, gives core states with only large gap associated Andreev bound states that are confined at a smaller length scale\cite{Sonier04}. The definition of the vortex core radius is somewhat arbitrary\cite{Kogan05} and depends on the particular experiment. It is found that, generally, the magnetic field produces a slight reduction of the coherence length $\xi$ in conventional superconductors \cite{Kogan05}. Measurements of vortex profiles at different magnetic fields support this idea\cite{Golubov94}. Nevertheless, it is not straightforward to relate magnetic field induced modifications of core size to the superconducting density of states \cite{Hayashi97}.

Our measurements provide a comparison between zero current and an applied current at the same magnetic field. We observe an increase in size of the vortex shape in the tunneling conductance map with an applied current. The radial position of the minimum in the slope of the radial dependence of the tunneling conductance is located at a slightly larger radius with an applied current, as compared to its value at zero current (Fig.5). We observe the creation of states in between vortices when smaller size gap features are further closed by the application of the current. Our measurements show that the actual core radius is a result of the full Fermi surface features, and that the electronic properties of different parts of the Fermi surface are interlinked to give the spatial core shape.

On the other hand, the increased amount of quasiparticle states observed with a small current suggests that macroscopic measurements could give variations as a function of current. Our result points out that  low currents can affect the gap structure of multigap or strongly anisotropic superconductors. Current should give additional directional information, providing new ways of probing anisotropic gap structures from macroscopic measurements\cite{Boaknin03,Bouquet01,Sologubenko02,Seyfarth08,Seyfarth05,Seyfarth06,Suderow98,Machida12}.

\begin{figure}[ht]
\includegraphics[width=8cm,keepaspectratio, clip]{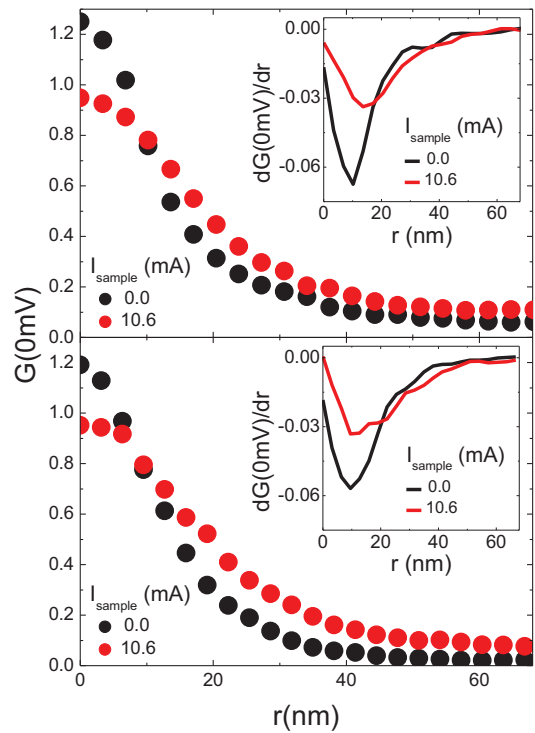}
\vskip -0cm \caption{Evolution of the normalized zero bias tunneling conductance $G_0$ at 200 mK and 0.1 T with the distance from the center of the vortex $r$ at zero current (black points) and at $I_{sample}$ = 10.6 mA (red points) along the direction of a ray (upper panel, Fig.2 upper panels) and along the one in between rays (lower panel, Fig.2 lower panels). The data has been averaged over the six equivalent directions in each case. The radial derivative of the tunneling conductance $dG_0/dr$ (numerically taken point by point from the data in the figure) is shown, for each case, in the insets (NC is normalized conductance).} \label{Fig5}
\end{figure}

In conclusion, we have used current driven scanning tunneling spectroscopy to observe that a small current affects the local density of states and the vortex bound states of 2H-NbSe$_{2}$. Superconductors with multiple gaps present sizeable variations of the gap structure as a function of the applied current.

We would like to thank discussions with I. Guillam\'on, A. Pautrat, A. Mel'nikov, A.I. Buzdin and V. Vinokur. P. Rodiere provided us with the 2H-NbSe$_{2}$ sample. We acknowledge support of Spanish MINECO and MEC (Consolider Ingenio Molecular Nanoscience CSD2007-00010 program, FIS2011-23488, ACI-2009-0905 and FPU grant), of COST MP2101, and of the Comunidad de Madrid through program Nanobiomagnet.


\end{document}